\pdfoutput=1
\documentclass{mem}
\usepackage{natbib}
\usepackage{txfonts}
\usepackage{balance}
\usepackage{graphicx}
\usepackage[a4paper]{hyperref}

\begin{document}
\title{A new code for radiation processes in high energy plasmas}
\author{R. \,Belmont\inst{1}  \and J. \, Malzac\inst{1} \and A. \, Marcowith\inst{2} }
\institute{Centre d'Etude Spatiale des Rayonnements, 9 rue du Colonel Roche, BP44346,  31028 Toulouse Cedex 4, France, \email{belmont@cesr.fr} \and Laboratoire de Physique Th\'eorique et Astroparticules, Universit\'e Montpellier II, place Eugène Bataillon, 34095 Montpellier, France
}
\authorrunning{Belmont}
\titlerunning{Radiation processes in high energy plasma}
\abstract{
Extreme objects such as X-ray binaries, AGN, or $\gamma$-ray bursters harbor high energy plasmas whose properties are not well understood yet. Not only are they responsible of the hard X- and $\gamma$-ray emission we observe but also they have a strong influence on the main dynamics and energetics of these objects themselves. 

Here we present a new kinetic code that solves the evolution equations for particles and photons around compact objects. It produces spectra that will be compared with observations from Simbol-X to constrain the radiation and acceleration processes in these objects. 
}
\maketitle{}

\section{Introduction}
Investigating the time evolution of interacting particles and photons implies to deal with integro-differential equations that cannot be solved analytically, and numerical simulations are required. 
We have started the development of a new code that deals with particles from sub- to ultra-relativistic regime, and includes interactions with the photon distribution such as self-absorbed synchrotron and Comptonization. It is time dependent and so will be able to address the variability observed in the sources light curves and spectra. 

\section{The code}
The code solves simultaneously two identical equations that describe the time evolution of both the distribution of particles in the the momentum space ($x=p/mc$) and that of photons in the energy space ($x=h\nu/mc^2$) :
$$ \partial_t N = \partial_x \left( A + \partial_x (D N)\right) + Q_{\rm inj} - \frac{N}{T_{\rm esc}} $$
The sink term $ N/T_{\rm esc}$ represents the escape of particles/photons from the system. The source term $Q_{\rm inj}$ includes the direct injection of particles/photons in the system (e.g. loading of matter or seed photons from an accretion disc) as well as interactions between the two populations (see hereafter). When possible/required, these interactions are treated in the Fokker-Planck approximation that leads to the first two terms corresponding to advection (A) (i.e. heating/cooling) and diffusion (D) in the momentum/energy space.

The distributions are discretized in bins. Since the Courant condition for explicit methods sets a very small time step when the energy range spans over several orders of magnitude, a semi-implicit method is used. The Chang-Cooper method \citep{CC70,PP96} is inaccurate when solving the FP equation with self-absorbed synchrotron for it leads to a poor energy conservation. Rather, we use a scheme based on the equations written above, which insures a number and energy conservation to machine precision.

\section{Radiation processes}
So far, two radiation processes have been implemented. \\
\textbf{Self-absorbed cyclo-synchrotron radiation:} \\
 The power spectrum of a single electron is tabulated from a combination of asymptotic expressions to get a good accuracy in all regimes \citep{GS91}. Contributions of the cyclo-synchrotron radiation to the FP coefficients in the equation for the particles are calculated from expressions given in \citet{GHS98}. \\
\textbf{Compton scattering:} \\
Including the Compton contribution for all energy regimes requires to compute the exact distribution $P(p_0,\omega_0;\omega)$ of photons resulting from the interaction of particles of momentum $p_0$ with isotropic photons of energy $\omega_0$. In the small angle scattering limit, the contribution to the Fokker-Planck coefficients is computed from the first moments of the scattered distribution. However, for the photon equation or for large angle scattering of particles, the FP approximation is not relevant and an exact integral treatment is used \citep[see][]{NM98}. 

\section{First results}
\begin{figure}
\includegraphics[scale=.38]{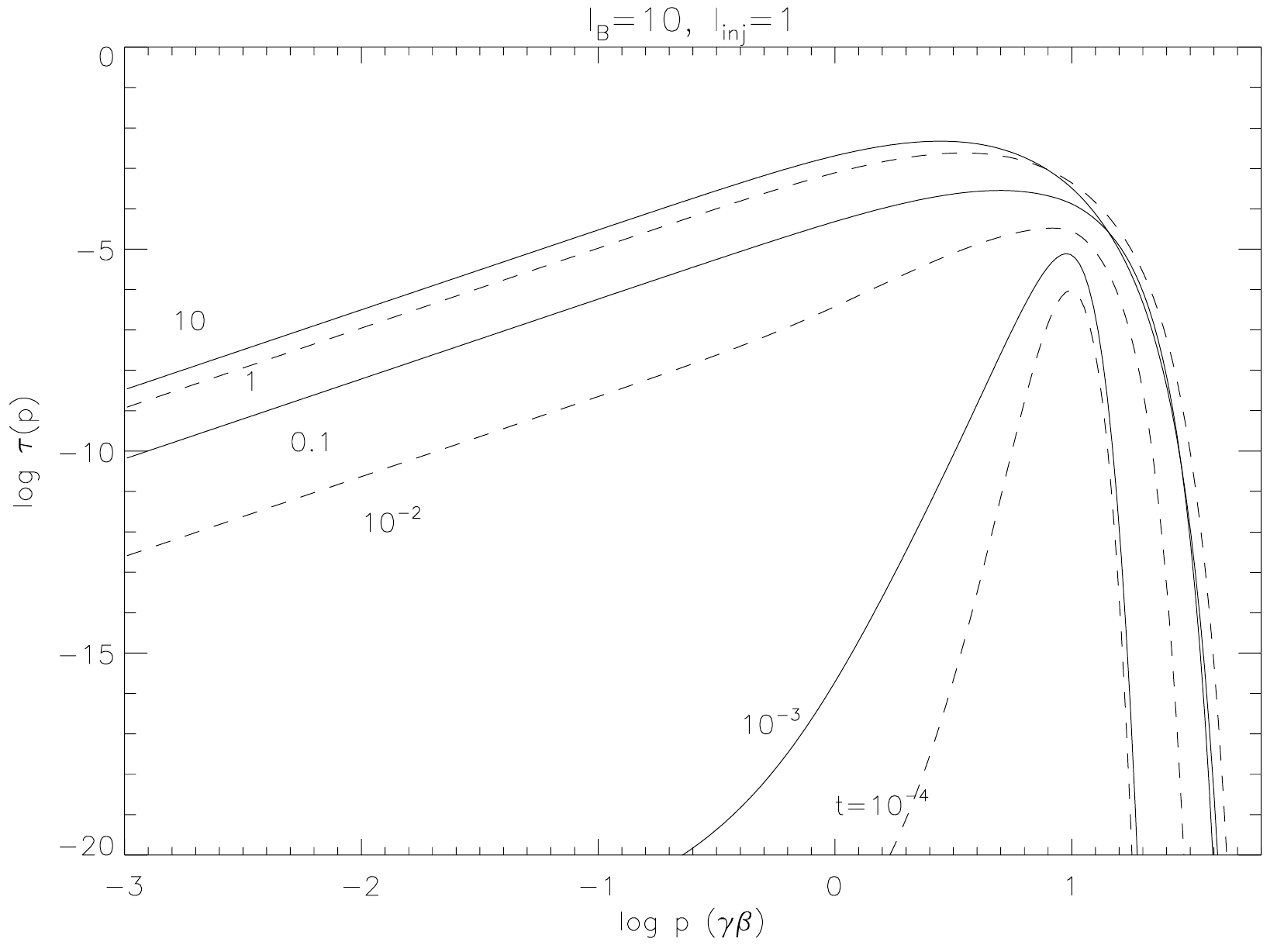} \label{fig}
\caption{Time evolution of the particle population when mono-energetic particles are constantly injected in an empty system (the time is normalized by the light crossing time of the system $R/c$ and the distribution by $R\sigma_T$).}
\end{figure}

Here we present a first use of the code. Several clues seem to indicate thermal plasmas in AGN and in X-ray binaries although acceleration processes rather produce power law distributions. Simple particle-particle collisions by Coulomb interactions are too rare to account for this thermalization. \citet{GGS88} suggested that exchange of energy between particles by exchanging synchrotron self-absorbed photons is much more efficient. Fig. \ref{fig} shows results very similar to those of \citet{GHS98}: as the the corona is filled by particles, their distribution evolves to a Maxwellian one on a few synchrotron times scales, showing that exchange of photons is a very efficient thermalization process.

\section{Conclusion}
We have presented the basic properties of the code at its present stage of development and one first result. New features will be added in order to give a complete modelling of high energy plasmas in microquasars, AGN, and $\gamma$-ray bursts... Among them: Coulomb interactions, non-absorbed cyclo-synchrotron coupled to other radiation processes, acceleration processes (e.g. Fermi), pair creation/anihilation...

\bibliographystyle{aa}

\end{document}